\documentstyle[prl,aps,epsfig,multicol]{revtex}

\onecolumn

\begin{document}
\title{Quantum state engineering on an optical transition and decoherence in a Paul trap}
\author{Ch. Roos, Th. Zeiger, H. Rohde, H. C. N\"agerl, J. Eschner,
D. Leibfried, F. Schmidt-Kaler, and R. Blatt}
\address{Institut f{\"u}r Experimentalphysik, Universit{\"a}t Innsbruck,\\
Technikerstra{\ss}e 25, A-6020 Innsbruck, Austria}
\date{\today}
\maketitle

\begin{abstract}
A single Ca$^+$ ion in a Paul trap has been cooled to the ground state of vibration with up to 99.9$\%$ probability. Starting from this Fock state $|n=0\rangle$ we have demonstrated coherent quantum state manipulation on an optical transition. Up to 30 Rabi oscillations within 1.4 ms have been observed. We find a similar number of Rabi oscillations after preparation of the ion in the $|n=1\rangle$ Fock state. The coherence of optical state manipulation is only limited by laser and ambient magnetic field fluctuations. Motional heating has been measured to be as low as one vibrational quantum in 190 ms.
\end{abstract}




\begin{multicols}{2}

Trapped and laser cooled ions in a Paul trap are currently considered to be promising candidates for a scalable implementation of quantum computation. Internal states of the ions serve to hold the quantum information (qubits) and an excitation of their common vibrational motion provides the coupling between qubits necessary for a quantum gate. The Cirac-Zoller proposal \cite{CiracZoller} requires that initially all ions are optically cooled to the ground state of motion and that the whole system can be coherently manipulated and controlled. This ideal situation is compromised by coupling to the environment in a real experiment. Therefore, experimental investigation of the limits of engineering a trapped ion's quantum state is an important step towards implementation of a quantum computer. In this letter, we present experiments on quantum state engineering with a single trapped $^{40}$Ca$^+$ ion that is initially prepared in the motional and electronic ground state. We manipulate the ion's motion using an optical transition to a metastable excited level. We measure the coherence time of this process as well as motional heating rates.

Ground state cooling has been achieved so far with a single $^{199}$Hg$^+$ ion \cite{Diedrich}, and with $^9$Be$^+$ \cite{Monroe}, using resolved sideband cooling on either a quadrupole or a Raman transition. With a cooling method similar to the Hg$^+$ experiment, we reach 99.9$\%$ of motional ground state occupation within 6.4~ms. We find a motional heating rate of one phonon in 190 ms, much smaller than in the trap used for the $^9$Be$^+$ experiment. 

The ion trap used in our experiment is a conventional 3D-quadrupole Paul trap \cite{Paul}. The ring electrode is made of 0.2 mm molybdenum wire and has an inner diameter of 1.4 mm. The endcaps are formed by two pieces of the same material that are rounded at the tips. The endcap to endcap distance is 1.2 mm. The radio-frequency drive field at 20.8 MHz is fed to the ring via a helical step-up circuit with a quality factor of 100. With a drive power between 0.5 W and 2.2 W we observe motional frequencies ($\omega_x$, $\omega_y$, $\omega_z$)/(2$\pi$) between (0.96, 0.92, 2.0) MHz and (2.16, 2.07, 4.51) MHz along the respective trap axes ($z$ denotes the axial direction of the trap, the degeneracy of $x$ and $y$ is lifted by small asymmetries of the setup).

\begin{center}
\begin{figure}[tbp]
\epsfxsize=0.6\hsize
\epsfbox{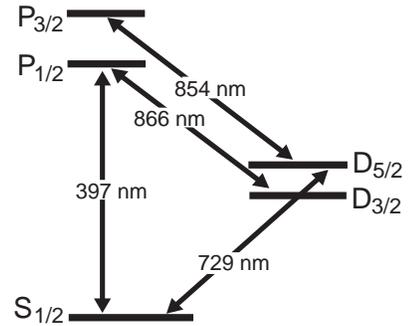} 
\caption{Relevant energy levels of $^{40}$Ca$^+$ and the corresponding transition wavelengths. \label{fig1}}
\end{figure}
\end{center}

$^{40}$Ca$^+$ is a hydrogen-like ion with one valence electron and no hyperfine structure. All relevant transitions are easily accessible by solid state or diode lasers (see Fig.~1) \cite{Naegerl}. In our experiment we Doppler-cool the ion and detect the internal state on the S$_{1/2}$ to P$_{1/2}$ transition at 397 nm, excited with a frequency-doubled Ti:Sapphire laser. This transition has a natural linewidth of 20 MHz and is not closed since the ion may decay to the metastable D$_{3/2}$ level with a branching ratio of 1:15. A diode laser at 866~nm serves to repump the ion via P$_{1/2}$ thus closing the cooling cycle. As the upper internal level for quantum state engineering and sub-Doppler cooling we use the metastable D$_{5/2}$ level with a natural lifetime of 1 s. The S$_{1/2}$ to D$_{5/2}$ quadrupole transition at 729~nm is excited with a highly stable Ti:Sapphire laser (bandwidth $<$ 1 kHz in 1 s averaging time). We apply a bias magnetic field to lift the degeneracy of sublevels in the ground and excited state manifolds. Prior to coherent manipulations the electronic ground state is prepared in pure S$_{1/2}(m=1/2)$ by optical pumping. We can detect whether a transition to D$_{5/2}$ occurred by applying the beams at 397 nm and 866 nm and monitoring the fluorescence of the ion. Typically, we collect 40 fluorescence photons on a stray light background of 2 photons in 2 ms if the ion is in the electronic ground state. This way the internal state of the ion is discriminated with an efficiency close to 100$\%$ \cite{shelving}. Another diode laser at 854 nm is used to repump the ion from the D$_{5/2}$ level to the electronic ground state via the P$_{3/2}$ level.

Calcium ions are loaded into the trap from a thermal beam by crossing it with an electron beam inside the trap volume. To make sure the ion rests at the node of the quadrupolar rf field, we compensate for stray DC-fields by applying small bias voltages to the endcaps and two nearby auxiliary electrodes.

Preparation of the motional ground state is accomplished by a two-stage cooling process. First the ion is cooled to the Doppler limit by driving the S$_{1/2}$ to P$_{1/2}$ dipole transition. In the second stage, a resolved-sideband cooling scheme similar to Ref.~\cite{Diedrich} is applied on the narrow S$_{1/2}$ to D$_{5/2}$ quadrupole transition: The laser is red detuned from the line center by the trap frequency $\omega$ (red sideband), thus removing one phonon with each electronic excitation. The cooling cycle is closed by a spontaneous decay to the ground state which conserves the phonon number. When the vibrational ground state is reached the ion decouples from the laser. The weak coupling between light and atom on a bare quadrupole transition would necessitate long cooling times. However, the cooling rate is greatly enhanced by (i) strongly saturating the transition and (ii) shortening the lifetime of the excited state via coupling to a dipole-allowed transition. 

For coherent spectroscopic investigation and state engineering on the S$_{1/2} \leftrightarrow$ D$_{5/2}$ transition at 729 nm we use a pulsed technique which consists of five consecutive steps. (i) Laser light at 397~nm, 866~nm, and 854~nm is used to pump the ion to the S$_{1/2}$ ground state and prepare the vibrational state at the Doppler limit $E=\hbar\Gamma/2$ \cite{Wineland79}. This corresponds to a mean vibrational quantum number $\langle n \rangle =10$ for $\omega=(2\pi)\:1$~MHz. (ii) Then the S$_{1/2}(m=+1/2)$ sub-state is prepared by optical pumping with $\sigma^+$ radiation at 397 nm. A magnetic field of 4 Gauss at 70$^\circ$ to the $k$-vector direction of the light at 729~nm provides a quantization axis and splits the 10 Zeeman components of the S$_{1/2} \leftrightarrow$ D$_{5/2}$ transition in frequency. (iii) Sideband cooling step: The S$_{1/2}(m=1/2) \leftrightarrow$ D$_{5/2}(m=5/2)$ transition is excited on one of the red sidebands at approximately 1~mW laser power focused to a waist size of 30~$\mu$m. The laser at 854~nm is switched on to broaden the D$_{5/2}$ level at a power level which is set for optimum cooling. Optical pumping to the S$_{1/2}(m=-1/2)$ level is prevented by interspersing short laser pulses of $\sigma^+$-polarized light at 397 nm. The duration of those pulses is kept at a minimum to prevent unwanted heating.
(iv) State engineering step: Then we excite the S$_{1/2}(m=+1/2) \leftrightarrow$ D$_{5/2}(m=+5/2)$ transition at 729 nm with laser pulses of well controlled frequency, power, and timing. These parameters are chosen according to the desired state manipulation. (v) Final state analysis: We collect the ion's fluorescence unter excitation with laser light at 397 nm and 866 nm and detect whether a transition to the shelving level D$_{5/2}$ has been previously induced. 

Sequence (i)-(v) is repeated typically 100 times to measure the D$_{5/2}$ state occupation $P_D$ after the state engineering step. We study the dependence of $P_D$ on the experimental parameters such as the detuning $\delta\omega$ of the light at 729 nm with respect to the ionic transition or the length of one of the excitation pulses in step (iv). The duration of a single sequence is typically 20 ms, so we can synchronize the sequence with the ac power line at 50 Hz to reduce ac-magnetic field fluctuations.

For the quantitative determination of the vibrational ground state occupation probability $p_{0}$ after sideband cooling we compare $P_D$ for excitation at $\delta\omega=-\omega$ and $\delta\omega=+\omega$, i.~e. on the red and the blue sideband. If the vibrational ground state is reached
with 100$\%$ probability, $P_D(-\omega_{trap})$ vanishes completely. For a thermal phonon probability distribution, the ratio of excitation on the red and blue sideband is given by $P_ {red}/P_ {blue} = \langle n \rangle / (1+ \langle n \rangle)$.

\begin{figure}[tbp]
\epsfxsize=0.95\hsize
\epsfbox{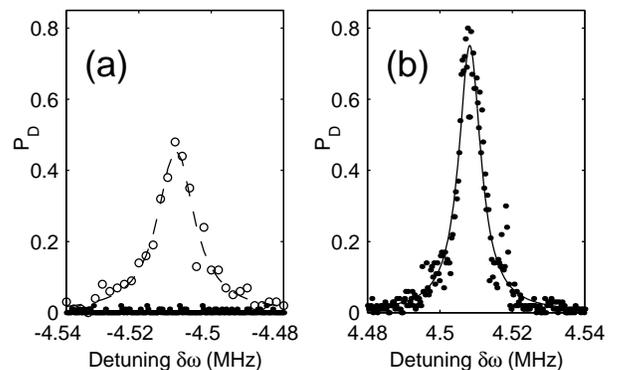} 
\caption{Sideband absorption spectrum on the S$_{1/2} (m=+1/2)$ $\leftrightarrow$ D$_{5/2} (m =+5/2)$ transition after sideband cooling (full circles). The frequency is centered around (a) red and (b) blue sideband at $\omega_z=4.51$~MHz. Open circles in (a) show the red sideband after Doppler cooling. Each data point represents 400 individual measurements. \label{fig2}}
\end{figure}

The ground state occupation after sideband cooling is determined by probing sideband absorption immediately after the cooling pulse. Fig.~2 shows $P_D(\omega)$ for frequencies centered around the red and blue $\omega_z$ sideband. Comparison of the sideband heights yields a 99.9$\%$ ground state occupation for the axial mode when $\omega_z = (2\pi)\:4.51$~MHz. By cooling the radial mode with $\omega_y = (2\pi)\:2$~MHz, we transfer 95$\%$ of the population to the motional ground state.
Ground state cooling is also possible at lower trap frequencies, however slightly less efficient. At trap frequencies of $\omega_z = (2\pi)\:2$~MHz, $\omega_y = (2\pi)\:0.92$~MHz we achieve $\langle n_z\rangle=0.95$ and $\langle n_y\rangle =0.85$, respectively. The x-direction is left uncooled because it is nearly perpendicular to the cooling beam. We also succeeded in simultaneously cooling all three vibrational modes by using a second cooling beam and alternating the tuning of the cooling beams between the different red sidebands repeatedly.

The best cooling results of 99.9$\%$ ground state occupation were achieved with a cooling pulse duration of $\tau_{cool}$ = 6.4 ms. The power of the cooling laser was set to about 1~mW which yielded the lowest value of $\langle n\rangle$ in the experiments. To study the cooling dynamics from the Doppler limit into the final state we varied $\tau_{cool}$ between zero and the maximum of 6.4 ms and determined the resulting ground state occupation. Fig.~3 shows $\langle n\rangle$ as a function of $\tau_{cool}$. We find that initially $\langle n\rangle$ decreases rapidly, then it tends to its finite final value. The decay constant, or cooling rate, determined from the data is 5~ms$^{-1}$. Both this value and the exponential behaviour are consistent with the expected 3-level dynamics during the sideband cooling process, taking into account our experimental parameters. The finite cooling limit is determined mainly by non-resonant excitation of the ion out of the ground state and heating in the subsequent spontaneous emission. The value is also overestimated, i.e. an upper limit, since no background was subtracted from the spectra.

\begin{figure}[tbp] 
\epsfxsize=0.95\hsize
\epsfbox{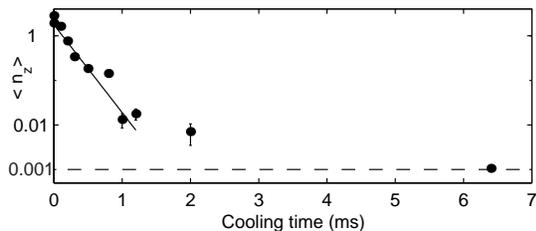} 
\caption{Cooling dynamics: Mean phonon number $\langle n_z \rangle$ {\it vs.} cooling pulse length deduced by sideband absorption measurements. The solid line assumes an initial exponential decay with 5~ms$^{-1}$ decay constant. The dashed line indicates our detection limit set by off-resonant excitation to the D level. \label{fig3}}
\end{figure}

Starting from the vibrational ground state, arbitrary quantum states can be created. To demonstrate coherent state engineering and investigate decoherence we excite, in step (iv), Rabi oscillations with the ion initially prepared in Fock states of its motion. Radiation at 729 nm is applied on the blue sideband transition $|$S$,n_z\rangle \leftrightarrow\: |$D$,n_z+1\rangle$ for a given interaction time $\tau$ and the excitation probability $P_D$ is measured as a function of $\tau$. The Rabi flopping behaviour allows to analyse the purity of the initial state and its decoherence \cite{HarocheMeekhof}. Fig.~4a shows $P_D(\tau)$ for the $|n=0\rangle$ state prepared by sideband cooling. Rabi oscillations at $\Omega_{01}=(2\pi)\:21 $~kHz are observed with high contrast indicating that coherence is maintained for times well above 1 ms. 
For the preparation of the Fock state $|n=1\rangle$, we start from $|$S$,n=0\rangle$, apply a $\pi$-pulse on the blue sideband and an optical pumping pulse at 854 nm to transfer the population from $|$D$,n=1\rangle$ to $|$S$, n=1\rangle$. Currently, the fidelity of this process is limited to about 0.9 by the non-ideal population transfer during the initial $\pi$-pulse (c.f.~the first oscillation in Fig.~4a) and the recoil heating of the optical pumping. As shown in Fig.~4b for the $|n=1\rangle$ initial state, we also observe high contrast Rabi oscillations, now at $\Omega_{12}=(2\pi)\:30$~kHz. The ratio of the Rabi frequencies ($\Omega_{n,n+1} \propto\sqrt{n+1}$ \cite{Blockley}) agrees with $\Omega_{01}/\Omega_{12}=1/\sqrt{2}$ within 1$\%$.
The Fourier transform of the flopping signals yields directly the occupation probabilities for the contributing Fock states $|n=0,1,2,3...\rangle$ \cite{HarocheMeekhof} and allows to calculate the purity of the prepared and manipulated states. For the 'vacuum' state $|n=0\rangle$, we obtain $ p_0 =0.89(1)$ with impurites of $ p_1 = 0.09(1)$ and $ p_{n\ge 2} \le 0.02(1)$. For the Fock state $|n=1\rangle$ the populations are $ p_0 =0.03(1)$, $ p_1 =0.87(1)$, $ p_2 =0.08(2)$, and $ p_{n \ge 3} \le 0.02(1)$. The measured transfer fidelity of about 0.9 agrees well with our expectation. Note that the Rabi flopping data were taken with less efficient cooling (lower trap frequency), and the number state occupation from the Fourier analysis is consistent with the temperature we determined by sideband measurements.

\end{multicols}

\begin{figure*}[tbp] 
\begin{center}
\epsfxsize=1\hsize
\epsfbox{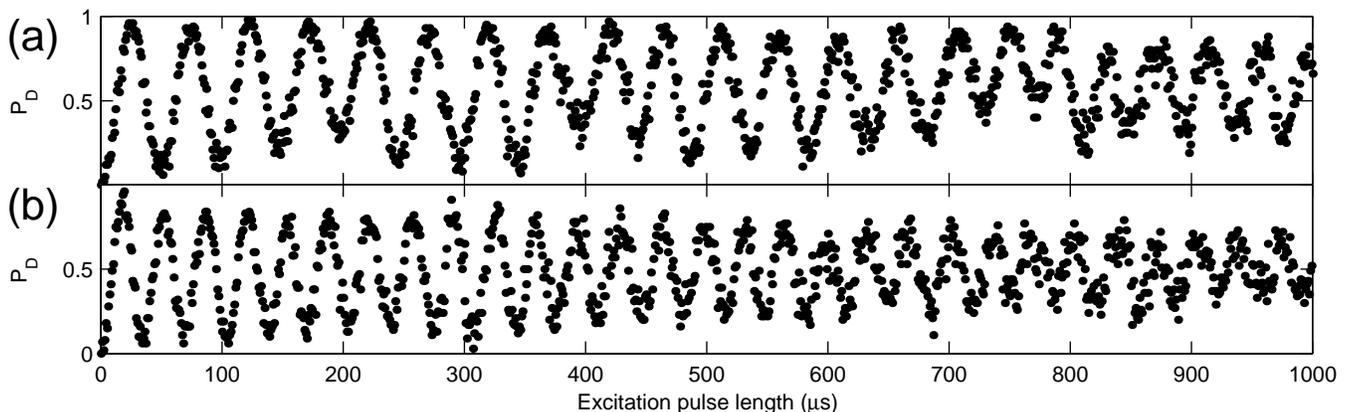} 
\caption{a) Rabi oscillations on the blue sideband for the initial state $|n=0\rangle$. Coherence is maintained for up to 1~ms. b) Rabi oscillations as in (a) but for an initial vibrational Fock state $|n=1\rangle$. \label{fig4}}
\end{center}
\end{figure*}

\begin{multicols}{2}

In order to investigate motional decoherence of the system, after sideband cooling the system is left alone for a certain delay time $t$ just interacting with the environment, i.e. with the surrounding electrodes and any possible perturbations acting on the motion of the ions. Then the resulting state is analyzed again by looking at the Rabi--flopping signal as described above. The result of such a measurement is shown in Fig.~5 where $\langle n_z \rangle$ is drawn as a function of the delay time $t$, yielding a heating rate of $d\langle n \rangle /dt = 0.0053$~ms$^{-1}$ (i.e. 1 phonon in 190 ms) at a trap frequency of $\omega_z=(2\pi)\:4$~MHz. For the radial y direction the heating rate was determined to be 1 phonon in 70 ms at $\omega_y = (2\pi)\:1.9$~MHz.

\begin{figure}[tbp]
\epsfxsize=0.95\hsize
\epsfbox{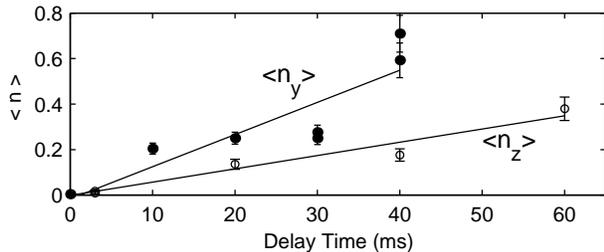} 
\caption{Heating rate measurements for the axial and radial vibrational modes at 4~MHz and 2~MHz, respectively. Heating rates are 1 phonon in 190~ms for the axial and 1 phonon in 70~ms for the radial mode. \label{fig5}}
\end{figure}

The contrast of the Rabi flopping decays to 0.5 after about 20 periods, therefore we expect that unitary manipulations equivalent to 40 $\pi$-pulses can be executed with a fidelity of 0.5 in our system. The observed decoherence time is consistent with independently measured values of the laser linewidth (below 1 kHz), the laser intensity fluctuations (below 3$\%$), and ambient magnetic field fluctuations (line shifts of $\pm$ 10 kHz at 50 Hz frequency).

In conclusion, we have engineered the quantum states of motion $|n=0,1 \rangle$ of a trapped ion that are relevant for quantum computation, using laser excitation on a forbidden optical transition. We have observed more than 30 periods of Rabi oscillations on the motional sidebands of this transition, thus showing that decoherence is negligible on the time scale of a few oscillations, i.e. in the time required for a quantum gate operation. We attribute the measured 1 ms decoherence time to laser and magnetic field fluctuations. Heating of the motional degrees of freedom has also been measured directly to happen at least 1 order of magnitude more slowly. This confirms that in the comparatively large trap that we use, heating is not a limiting process. 

The ability to engineer quantum states as well as the long coherence times and small heating rates that we find are fundamental requirements for quantum information processing with trapped ions. In a next step, these techniques will be applied to more than one trapped ion in a linear ion trap \cite{Naegerl}, which has similar dimensions and characteristics as the spherical trap used here. Ongoing improvements to laser and magnetic field stability will extend the coherence time beyond 1~ms. Another essential step, the individual addressing of ions in a linear trap with a laser beam, has already been demonstrated in that trap \cite{addressing}. In summary, multiple coherent gate operations with trapped ions and the scaling of the trapping and cooling technique to several ions seem well within experimental reach.

This work is supported by the Austrian "Fonds zur F\"orderung der wissenschaftlichen Forschung" within the project SFB15, and in parts by the European Commission within the TMR networks "Quantum Information" (ERB-FMRX-CT96-0087) and "Quantum Structures" (ERB-FMRX-CT96-0077).

\end{multicols}

\end{document}